\documentclass[manuscript]{aastex}
\usepackage{amsmath}
\usepackage{natbib}

\slugcomment{Submitted to ApJ}

\shorttitle{2nd order moment as observable in lensed quasars}
\shortauthors{Guerras et al.}

\begin{document}

\title{A Second-Order Moment of Microlensing Variability as a Novel Tool in Extragalactic Research}

\author{Eduardo Guerras\altaffilmark{1}, Xinyu Dai\altaffilmark{1}, Evencio Mediavilla\altaffilmark{2}}

\altaffiltext{1}{Homer L. Dodge Department of Physics and Astronomy, The University of Oklahoma,
Norman, OK 73019, USA}
\email{e.guerras.valera@gmail.com}

\altaffiltext{2}{Instituto de Astrof\'isica de Canarias, C/V\'ia L\'actea s/n, Sta Cruz de Tenerife 38205, Spain}


\begin{abstract}

We define a second-order moment of the observational differential microlensing curves that can be used to impose constraints on physical properties of lensed quasars. We show that this quantity is sensitive both to variations in the source size and the deflector mass. We formulize a methodology to recover the source size from the observational measurements when the mass spectrum is fixed. As a case study, we test it with a sample of four quadruple lenses, both in simulated scenarios and with real data from the \textit{Chandra X-ray Observatory}. In our simulations with a uniform stellar population the method works best to detect sources around $0.1$ Einstein radii, giving correct upper/lower limits for much smaller/bigger sizes without requiring a big leap in additional computational effort as compared to a single-epoch approach, yet taking advantage of multi-epoch information. We apply the method to a small sample of X-ray data from four objects assuming a range of star masses, and obtain a degeneracy relation between the source size and deflector mass. Combined with previous estimates for the size of the X-ray corona, the degeneracy relation suggests that X-ray microlensing is mainly induced by planetary mass objects.

\end{abstract}
    

    
\section{Introduction}

Quasar microlensing studies rely on the flux ratio anomalies induced by micro lenses, i.e. in what amount they depart from the predicted flux ratios in the absence of point deflectors embedded in the mass distribution of the lens. Temporal series of such flux ratios show uncorrelated variations between different lensed images that can be explained by discrete objects in the galactic halo. Direct imaging of the micro-lenses is impossible because of the cosmological distances involved. However it is possible to test different models by comparisons with observations. This requires a model distribution of micro-lenses and the generation of artificial flux ratios, either isolated or as a temporal sequence (light curves). Statistical fitting of the simulated light curves with the observational data \citep{Kochanek2004}  has proven to be a successful method for individual lens systems \citep[e.g.,][]{morgan2008,Morgan2012,Dai2010,Mosquera2013,Blackburne2014,Blackburne2015,Macleod2015} though, as new observations are appended to the archival data, the computational effort grows exponentially.

In many cases there is only one observation available per object, especially for spectral or polarimetry observations, and a single epoch microlensing analysys approach is utilized \citep[e.g.,][]{Fian2018}. The microlensing constraints from these single-epoch analysis are reliable when using joint samples, where objects are grouped either under the assumption of the physical similarity of the emitting source, or after some rescaling (e.g. in gravitational radii of the central SMBH). This single-epoch approach cannot achieve the precision of track-fitting, but there are situations that fits well for single epoch studies. First, they require less assumptions with regard to the physical properties of the system, e.g. it can be argued that a single-epoch approach does not rely significantly on assumptions about the mass of the individual deflectors, except for special combinations of values of the macrolens parameters $\kappa$ and $\gamma$ and a wide mass spectrum \citep{Schechter2004}. Second, a single-epoch treatment is computationally cheap. Both characteristics make them useful for certain tasks, e.g., exploring a wide parameter space in the physical properties of lens halos or characterizing a group of lensed quasars.

However, the addition of observations to single-epoch studies poses additional challenges. This is not the case if the time elapsed between observations is much shorter than the typical scale for microlensing variations, i.e. the time for the source to cross the Einstein radius\footnote{The typical source size scale for microlensing is given by the Einstein angle:
\begin{equation} \label{eq:AE}
 \theta_{E}=\sqrt{\frac{4GM}{c^{2}}\frac{D_{ds}}{D_{d}D_{s}}}.
\end{equation}
where $M$ denotes the deflector mass and $D_{ol}$, $D_{os}$, and $D_{ls}$ are the angular diameter distances between the observer, lens, and source respectively \citep[e.g.,][]{Wambsganss2006}.
} because then the data should not be considered as statistical independent and averaging can be justified. Conversely, if data are separated by a long enough interval \citep[see table of typical Einstein radius crossing times in][]{MosqueraKochanek2011}, then the ratios from different epochs should be considered independent and treated as additional single-epoch data points in the joint sample. The problem arises with situations in between, where choosing one or the other strategy requires an arbitrary decision.

Several studies have faced this challenge, and intermediate solutions were used to take advantage of multi-epoch observations without performing a full track-fitting analysis study like, e.g. breaking light curves in intervals that are averaged separately and treated like sinle-epoch observations \citep{Munoz2016} or, as we did in a previous study \citep{Guerras2017}, comparing the averages of the light curves against numerical models of such averages. The later strategy solved the problem of ignoring the information loss when averages are treated as if they were single-epoch data, however it cannot be a satisfactory solution in the long term because as we add more data to the archival observations, the information may get diluted in the averages.

The standard deviation of the microlensing ratios provides another observable, since we also found in \cite{Guerras2017} the observational fact that the standard deviation or RMS of the microlensing flux ratios has a functional dependency with the average of the light curves, hence it also carries redundant information on physical properties of the system. But we are not interested here in the RMS of the flux ratios, but rather in the RMS of the logarithmic flux ratios, i.e., magnitudes, as we show in Section \ref{sec:2nd_section} that the RMS of magnitude differences may carry independent information.\footnote{When working with a single observation is rather indistinct to use the ratio of linear fluxes, or the difference in magnitude scale, but that is not the case when dealing with RMS of the light curves for elementary arithmetic reasons.} Because of the degeneracy between amplitude and global multiplying factor of any curve when converted to log scale, the RMS of the logarithmic microlensing ratios is a different observable, and some studies \citep{Dai2010, Chen2011, Chen2012} have already pointed at high values of the RMS of the logarithmic light curves as evidence for the presence of microlensing in lensed quasars. The standard RMS formula quantifies the dispersion around the average of the sample; however the flux ratio light curve can be shifted as a whole by microlensing, specially for campaigns that are not very long as compared with the typical time scales for microlensing variability, and the microlensing magnification offset information would be lost.

We instead define a quadratic quantity similar to the standard deviation, computing the dispersion around the baseline flux ratio (the value in complete absence of microlensing) instead of the average of the flux ratio light curve. This quantity will be affected both by fluctuations around an average value and by the shift of that average value with respect to the no-microlensing baseline. We will explore the utility of this quantity under the context of extragalactic microlensing, and propose a method to extract physical constraints from it.

In Section \ref{sec:1st_section} we stablish the definitions, describe the numerical simulations set up, and the observational data. 
In Section \ref{sec:2nd_section} we explore the behavior of the new quantity under variations of the source size and the length of the observational campaign, and its correlation with other observables.
In Section \ref{sec:3rd_section} we propose a temptative methodology to recover the physical information encoded by this observable.
Section \ref{sec:4th_section} applies the method to a sample of four lensed quasars observed by \textit{Chandra} in the X-ray band.
We assume a flat $\Lambda$CDM cosmology of $H_0 = 72~{\rm km~s^{-1} Mpc}^{-1}$, $\Omega_m = 0.3$, and $\Omega_\Lambda=0.7$.


\section{Microlensing Simulations, Observables and definitions}\label{sec:1st_section}

\subsection{Magnification Patterns and Observational Data}

To simulate microlensing, we initially model the microlens population as a uniform distribution of point deflectors randomly scattered across the lens plane. The choice of a particular mass is not relevant as long as all the projected transversal distances, including the source size and the track length, are measured in Einstein radii, which we use in this Section and in Sec. \ref{sec:2nd_section} and \ref{sec:3rd_section} as well, to develope a methodology to work with the new quantity.

Once the deflector mass function is fixed, each possible magnification pattern in the source plane is identified by three values $\kappa_{*}/\kappa, \kappa, \gamma$, mass fraction in deflectors, local surface density, and shear respectively. For the systems used in this study the values of these parameters are taken from the models used in \cite{Guerras2017} (see references therein) and are summarized in Table \ref{tab:kappasgammas}. We then use the Inverse Polygon Mapping algorithm \citep{Mediavilla2006,Mediavilla2011b} to compute sets of magnification maps with a dimension of $8000^2$ pixels spanning a physical size corresponding to $32.0$ Einstein radii ($0.004$ $E_r$ per pixel). This size is consistently maintained throughout Sections \ref{sec:2nd_section} to \ref{sec:3rd_section}, and to model the microlensing effect on extended sources in these Sections we convolve these patterns with a Gaussian kernel $I(r_g) \propto \exp[-r_g^2/(2r_{s}^{2})]$, for a set of source sizes $r_{s}/0.004=e^{0.08n}$ with $n=0,1, 2, ...$, from $0.004$ Einstein radii all the way up to $602$ Einstein radii, to make sure we have enough resolution at all size scales.

After we develope the necessary tools, in Sec. \ref{sec:4th_section} we apply them to real data, and for that we will use different mass values spanning the interval from $0.01$ $M_{\odot}$ to $1.0$ $M_{\odot}$, and the scaling will be in gravitational radii of the central SMBH. We will put to the test the same observational data from the \textit{Chandra X-ray Observatory} archive we analyzed in \cite{Guerras2017} for the lenses SDSS 0924+0219, HE 0435$-$1223, SDSS 1004+4112, and Q 2237+0305. The campaigns contain $6$, $10$, $11$, and $30$ observations spanning $5.6$, $7.3$, $9.4$, and $13.6$ years respectively. See Table \ref{tab:summaryofdata} for details, and the original paper for further details.

\subsection{Definitions}

When flux measurements are given in magnitude scale, the usual definition for differential microlensing $\Delta m_{BA}$ between components $A$ and $B$ of a lensed quasar can be written as:
\begin{equation}\label{eq:definition_microlensing}
m_{B} - m_{A} = (m^{1}_{B} - m^{1}_{A}) - (m^{0}_{B} - m^{0}_{A}) ,
\end{equation}
where $m^{1}_{i}$ is the flux of image $i$ in magnitude scale, and $m^{0}_{i}$ can be either the baseline flux in the absence of microlensing or the magnification produced by the smooth distribution of mass in the lens galaxy. The difference $(m^{0}_{B} - m^{0}_{A})$ can be estimated from the model, or in certain cases it can be measured by the IR or extended radio emission, and we will subtract it from both the models and the observational data. One convenient consequence that follows from Def. \ref{eq:definition_microlensing} is that the absence of microlensing is represented by $m_{B} - m_{A}=0$.

The RMS or standard deviation of $\Delta m_{BA}$ is:
\begin{equation}\label{eq:definition_standard_deviation}
\sqrt{\frac{1}{N}    \sum_{k}    \Big ( (m_{B,k}-m_{A,k}) - \langle m_{B}-m_{A} \rangle  \Big )  ^{2}} ,
\end{equation}
the summation being over all epochs in the light curve. This is the classical definition of second order moment of a distribution, that quantifies the average quadratic distance of the data to their average value $\langle m_{B}-m_{A} \rangle$ . However we are interested in quantifying the departure from the baseline value given by the flux ratios in the absence of microlensing, i.e. from a zero value which we will use instead of $\langle m_{B}-m_{A} \rangle$. Hence we define the quantity:
\begin{equation}\label{eq:definition_r}
\mathcal{R} = \sqrt{\frac{1}{N} \sum_{k} (m_{B,k}-m_{A,k})^{2}} ,
\end{equation}
or the mean quadratic distance of microlensing ratios in magnitude scale to the value of no-microlensing, the summation being over all epochs in the light curve. The mathematical treatment of this quantity will be similar to a usual standard deviation or RMS, but it has the advantage of quantifying both the oscillations and the global offset of the signal with respect to the baseline value where microlensing is absent, which may be relevant in campaigns that are not much longer than the time scale for microlensing.


\section{$\mathcal{R}$ is sensitive to variations in physical parameters.}\label{sec:2nd_section}

\begin{figure}
\epsscale{1.00}
\plotone{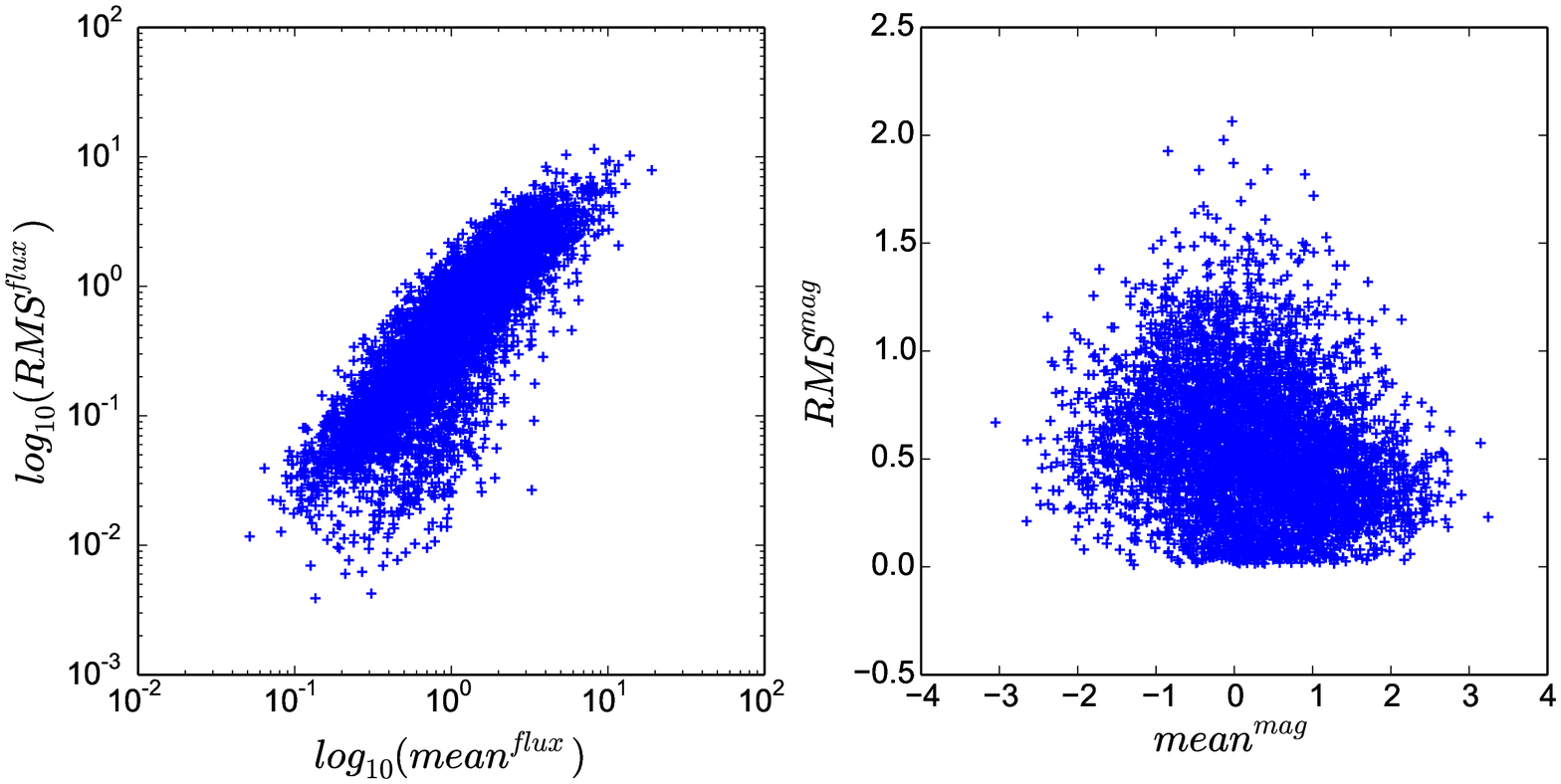}
\caption{The results of 5000 simulated campaigns of 30 observations each, along 0.5 Einstein radii tracks in random positions and orientations, for the microlensing C-A of Q 2237+0305. The standard deviation of the flux ratios shows the same dependency with the mean of flux ratios observed in the fiducial data (left panel), whereas the RMS of the magnitude differences vs the average of that difference shows no correlation (right panel). \label{fig:rms_vs_mean_simulation}}
\end{figure}

It can be shown experimentally \citep{Guerras2017} that a functional dependence exists between the RMS of microlensing flux ratios and the average of those ratios taken across observational campaigns. However, if the data are taken in magnitudes in the first place before computing averages and RMS as we define in Sec. \ref{sec:2nd_section}, the functional dependence does not hold in general. We can confirm this experimental observation with simulated data. Fig. \ref{fig:rms_vs_mean_simulation} shows these quantities extracted from $5000$ simulated campaigns of $30$ observations spanning $0.5$ Einstein radii each, for the microlensing of $C-A$ in Q 2237+0305 and a source size of $1$ light day for this system. The left pannel shows a linear correlation between the RMS of microlensing ratios and their average after they are expressed in the logarithmic space. However if the logarithm is taken on the microlensing ratios \textit{before} computing the statistics, the result shows randomly scattered points (right pannel).

This suggests that the RMS of the logarithmic quantities is not trivially related with the average, and similarly $\mathcal{R}$ should not be either. Hence, if $\mathcal{R}$ is sensitive to variations in physical parameters, it is in its own way. We expect that $\mathcal{R}$ as a new microlensing observable can be complementary to other observables or even carry physical information in situations where the other quantities are not appropriate, e.g for long observational campaigns where averaging the data may lead to information loss, and to explore this point we use the set of patterns described in Sec. \ref{sec:1st_section} to simulate $\mathcal{R}$ and perform some simple tests on the image pair $C-A$ in Q 2237+0305 to show that this quantity can be sensitive to variations in physical parameters of the system. The phase space of microlensing simulations can be huge \citep[e.g.,][]{Vernardos2013} and we have chosen Q 2237+0305 simply as an example for this preliminary exploration for several reasons. First, it allows us to perform the study in absence of time delays $< 1$ day \citep{Dai2003}. Second, images are at microlensing high optical depth, making microlensing phenomena more likely. Also, this very famous object has been studied for very long, so we have a relative long campaign in the \textit{Chandra} archive.

\begin{figure}
\epsscale{1.00}
\plotone{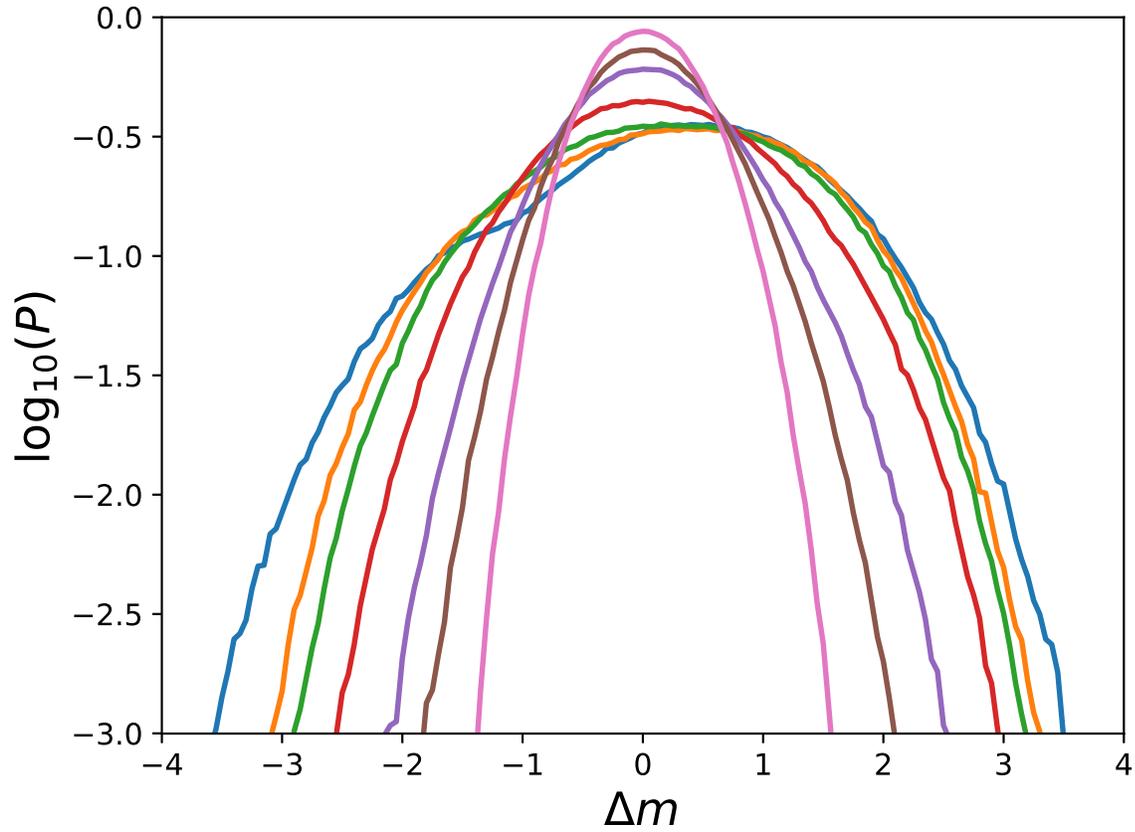}
\caption{Probability distribution of single-epoch differential microlensing for several source sizes. Each distribution is obtained from $10^7$ simulated observation pairs for the components C and A of Q 2237+0305. The source sizes range from 1 (wider, blue curve) to 140 light-days (innermost curve). \label{fig:histogs_singleepoch}}
\end{figure}

A single differential microlensing ratio $\Delta m_{CA}$ as defined in Eq. \ref{eq:definition_microlensing} can be simulated from a pair of measurements at random points of two magnification maps corresponding to images $C$ and $A$ respectively for a specific source size. We build a probability function from a large number of such random pairs. This probability follows a different distribution for each source size, in response to the known fact that microlensing is less sensitive the more spatially extended the source. Figure \ref{fig:histogs_singleepoch} shows the result of simulating such distributions for the differential microlensing $(C-A)$ of Q 2237+0305 for source sizes from $1$ light-day up to $140$ light-day, each distribution made from $10^7$ trials. These distributions are useful as a tool to understand what can be expected from a single epoch measurement. Fig. \ref{fig:histogs_singleepoch} shows that the compact X-ray emitting regions are likely to produce values in a big range, whereas the detection of microlensing of the BLR is more challenging because the values will be most often be closer to zero. The symmetry of Fig. \ref{fig:histogs_singleepoch} also tells us that we can also expect demagnification of image C with respect to image A due to microlensing magnification being stronger in $A$ with a significant probability, although positive values are in this case slightly more likely. 

\begin{figure}
\epsscale{1.00}
\plotone{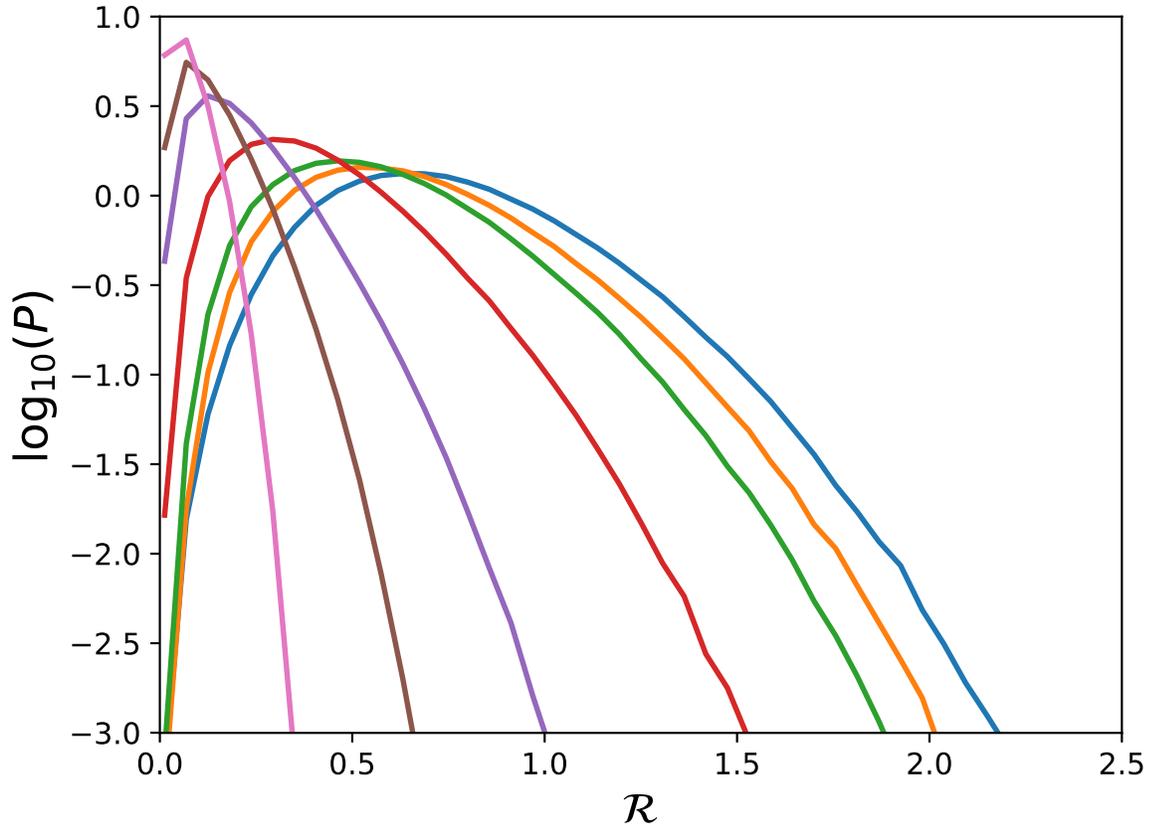}
\caption{Probability distribution of the quantity $\mathcal{R}$ computed from the differential microlensing C-A of Q 2237+0305 for source sizes ranging from $1$ light-day (blue) to $140$ light-days in size. Each distribution is obtained from $10^7$ simulated observation campaigns of 30 observations each, spanning $0.5$ Einstein radii, for the pairs for the components C and A of Q 2237+0305. \label{fig:histogs_rms63}}
\end{figure}

To get the probability distributions of $\mathcal{R}$ for the image pair $(C-A)$ of Q 2237+0305, instead of isolated point measurements on the maps we get sets of 30 data points on straight lines that expand $0.5$ Einstein radii on random positions and orientations on the same set of maps, convolved with the same set of sources ranging from $1$ light-day up to $140$ light-day. For each pair of tracks we compute $\mathcal{R}$ as defined in Eq. \ref{eq:definition_r}, and this is done $10^7$ times for each source size. As a result, (Fig. \ref{fig:histogs_rms63}) is different from its analogous for single-epoch microlensing, but most important, it shows a clear dependence of the probability of $\mathcal{R}$ with respect to the source size, being smaller sources more likely to yield high numbers. Here we rely on the assumption that it is possible to use a static map because the number of caustic crossing averaged over all directions should be the same in a pattern where caustics move and evolve with time or, at most, that we could do a equivalence by doing some correction to the length of the simulated tracks, which is out of the scope of this preliminary study.

\begin{figure}
\epsscale{1.00}
\plotone{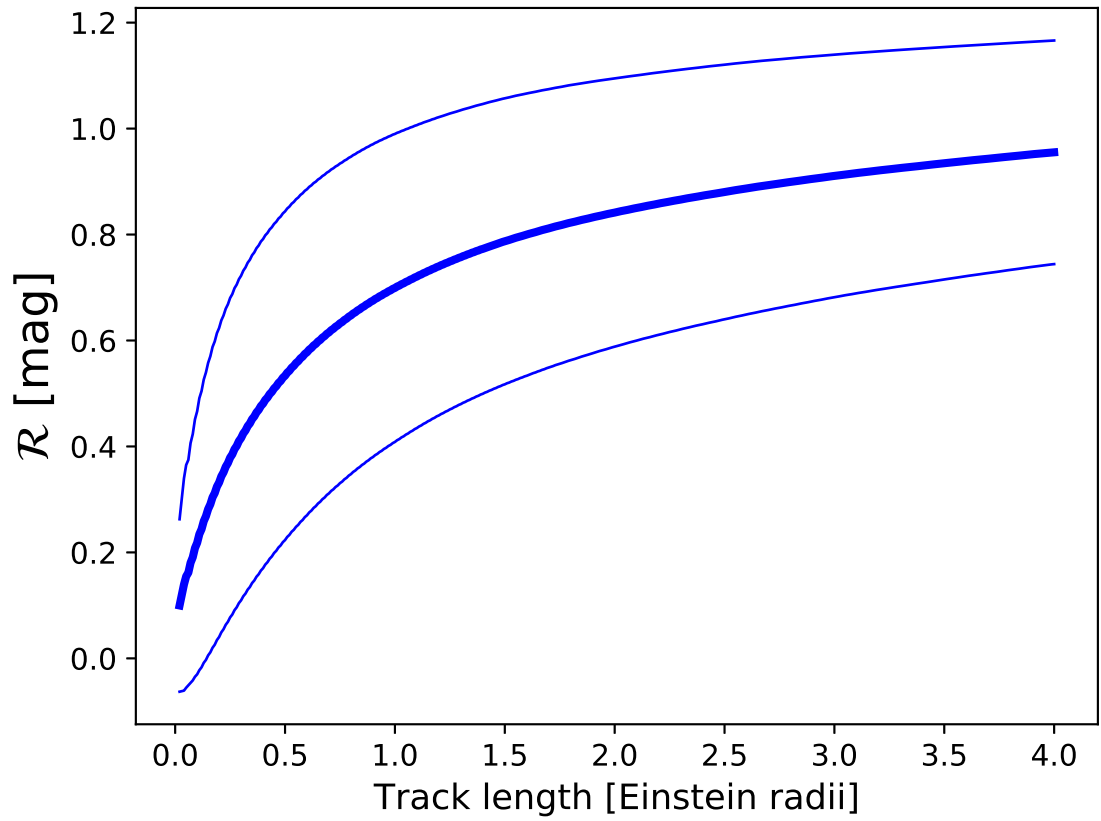}
\caption{The expected values of $\mathcal{R}$ for the differential microlensing C-A of Q 2237+0305 vs track length in Einstein radii, for a N=13 source size. The thin lines represent the 1-sigma limits. \label{fig:rms_vs_track_length}}
\end{figure}

The quantity $\mathcal{R}$ can be thought of as a measure of the variability when the background image crosses the caustic network of the magnification patterns. The length of the campaign may have some impact, and to explore this point we have simulated a set of tracks on the maps departing from two observations $0.01$ Einstein radii apart, adding observations until reaching $4$ Einstein radii long tracks. Fig. \ref{fig:rms_vs_track_length} shows the result of averaging the $\mathcal{R}$ values obtained in $10^6$ trials per track length. This figure suggest that the signal gets stronger for longer campaigns instead of getting diluted, as would be the case with the average. The length of a track against a static caustics background is related to the number of caustic crossing, hence a case can be made that $\mathcal{R}$ is useful to impose constraints on the relative velocity of the source if a fixed emitting region size is assumed or known by other means.


\section{A Practical Method}\label{sec:3rd_section}

As an example of the possibilities of the quantity we have defined, we propose a simple method to recover the source size, at a fixed campaign length against static caustic patterns. To find such method, we describe and review a single-epoch approach first, to look for a reasonable generalization.

\subsection{The single-epoch approach revisited.}

A single epoch microlensing ratio $\Delta m_{\beta \alpha}^{obs}$ can be compared with the prediction $\Delta m_{\beta \alpha}^{model}$ from a numerical model by means of the probability distributions generated from the magnification patterns. for that purpose, we can compute the $\chi^2$ statistic of the difference:

\begin{equation}\label{eq:chi2_single_epoch}
\chi^2=\sum_{\alpha} \sum_{\beta < \alpha} \Big ( \frac{ \Delta m^{obs}_{\beta \alpha} - \Delta m_{\beta \alpha} }{\sigma_{\beta \alpha}} \Big )^{2} ,
\end{equation}
where $\alpha$ $\beta$ go for $1$ to $4$ in the case of a quadruple object. The errors $\sigma_{\alpha \beta}$ are computed according to Equation $(7)$ in \cite{Kochanek2004}.

To get an estimate of the likelihood of the model, this needs to be done for a high number $N$ of trials in which $\Delta m^{obs}$ remains the same and $\Delta m^{model}$ adopts random values following the probability distribution given by the numerical model (i.e. the histograms of paired magnification patterns):
\begin{equation}\label{eq:average_of_trial_likelihoods}
L(r_s) \propto \sum^{N}\exp{\Big( -\frac{1}{2}\chi^2 \Big)}.
\end{equation}

This process is repeated with numerical models generated from different values of the physical parameter we want so study, in this case the source size $r_s$, and this yields the a priori probability distribution $L(r_s)$ for the parameter $r_s$. The process is quite erratic for a single object and epoch, but usually a number of lenses take part in the study, under the assumption of a universal validity of a quasar model, sometimes after a certain physical rescaling, e.g. by working in units of gravitational radii of the central SMBH \citep[e.g.][]{Jorge2015}. By using one observation per lens the statistical independence is granted and the resulting joint probability distribution is a multiplication of the individual distributions:
\begin{equation}\label{eq:definelikelihood} 
 P(r_{s}) \propto \prod_{n} L_{n} (r_{s}).
\end{equation}
where $n$ distinguishes between objects. Some examples can be found in \cite{Fian2018, Guerras2017, Munoz2016, Mediavilla2011a, Guerras2013a, Guerras2013b}.

We are restricting our study in this section to a sample with $4$ quasars with $4$ images each: Q 2237+0305, HE 0435$-$1223, SDSS 1004+4112, and SDSS 0924+0219, using the macrolens model parameters and local stellar mass fractions in \cite{Guerras2017} (see references therein). The values are summed up in table \ref{tab:kappasgammas}.

\begin{figure}
\epsscale{1.00}
\plotone{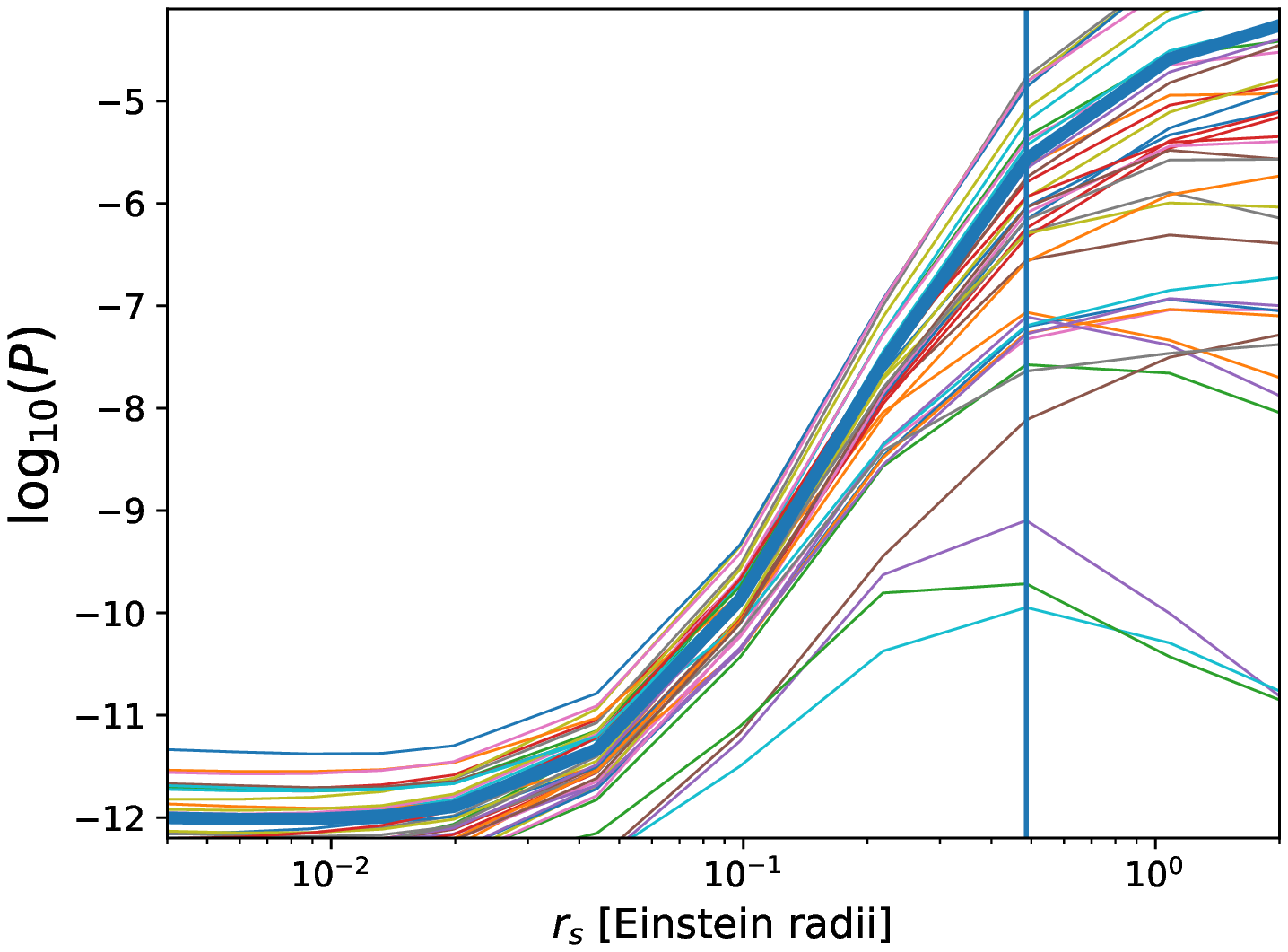}
\caption{Single-epoch joint-likelihood functions (not normalized) for $40$ random single-epoch observations simulated from magnification patterns corresponding to a source size of $0.5$ Einsteinradii (vertical blue line). The thick, blue curve represents an average. \label{fig:detail_single_big}}
\end{figure}

We have simulated $200$ single-epoch observations per image ($800$ measurements per object) separately for gaussian source sizes of sigma $0.5$, $0.1$, and $0.01$ Einstein radii. The simulated observations come from a different set of patterns where the random seed for the star positions generator was different than the one used for the generation of the models. The models account for a set of source sizes ranging from $0.004$ up to $602$ Einstein radii to make sure we have enough resolution at all scales. We have applied the discussed methodology to obtain $200$ joint probability distributions for the source size in the three cases, to see if we can recover the correct size from them. Figure \ref{fig:detail_single_big} shows that nearly all the realizations work well to give us a lower bound of the source size in the case of a $ 0.5 $ Einstein radii source. However, for the intermediate-sized source most predictions are too large and the expected values are widely dispersed. In the case of the smallest source, the predicted values ​​are erroneous by more than one order of magnitude and the dispersion is very large. The average curve predicts an erroneous value to more than two sigmas of the real one. The conclusion is that a set of only four lenses is too small for this single-epoch approach to have a reasonable predictive value, and that is why most single-epoch studies rely on bigger lens samples.

\subsection{A generalization for the study of $\mathcal{R}$}\label{subsec:generalization}

We are now generalizing the same steps to find a methodology we can apply in the case of the quantity we have defined. Eq. \ref{eq:chi2_single_epoch} resembles a $\chi^2$ statistics in the absence of covariance between the random variables, given that the experimental error is $\sigma_{\beta \alpha}$ \footnote{Other considerations are taken into account in the calculation of $\sigma_{\beta \alpha}$, see \cite{Kochanek2004} for details}. So we could think of an analogous expression to start with:
\begin{equation}\label{eq:chi2_adapted}
\chi^2=\sum_{\alpha} \sum_{\beta < \alpha} \Big ( \frac{ \mathcal{R}^{obs}_{\beta \alpha} - \mathcal{R}_{\beta \alpha} }{f_{\beta \alpha}} \Big )^{2} ,
\end{equation}
where again, $\alpha$ $\beta$ take values from $1$ to $4$ in the case of a quadruple object. However the meaning of $f_{\alpha \beta}$ is not trivial here because there is correlation between the random variables. The variability in one of the images will have an impact in $\mathcal{R}$ of any pair it takes part in, hence the quantities $\mathcal{R}_{ij}$ and $\mathcal{R}_{ik}$ that have image $i$ in common are no longer independent.

The absence of covariance in the case of single-epoch studies is due to the fact that, when dealing with differential microlensing measurements, we are only interested in the relative differences between the components of each pair.  To arrive to an analogous of Eq. \ref{eq:chi2_single_epoch} we should depart from the expression of a $\chi^2$ statistic in the general case with a variance-covariance matrix. This expression can be written as the quadratic form:
\begin{equation}\label{eq:chi2_six_dimensional}
\chi^2= \sum_{i=1}^{6} \sum_{j=1}^{6} \Big ( \mathcal{R}^{obs}_{i} - \mathcal{R}^{model}_{i} \Big ) \Big [ V^{-1} \Big]_{ij} \Big ( \mathcal{R}^{obs}_{j} - \mathcal{R}^{model}_{j} \Big ),
\end{equation}

Note that this is a general $\chi^2$ statistics of a six-dimensional random variable. Indexes have been remapped to match the six possible combinations of images:

$$(\beta = 1, \alpha = 2)\rightarrow i = 1 $$
$$(\beta = 1, \alpha = 3)\rightarrow i = 2 $$
$$(\beta = 1, \alpha = 4)\rightarrow i = 3 $$
$$(\beta = 2, \alpha = 3)\rightarrow i = 4 $$
$$(\beta = 2, \alpha = 4)\rightarrow i = 5 $$
$$(\beta = 3, \alpha = 4)\rightarrow i = 6 $$
so that
\begin{equation}\label{eq:definition_rms_with_index_remapping}
\mathcal{R}_i = \mathcal{R}_{\beta \alpha} = \sqrt{\frac{1}{N} \sum_k \Big ( m_{\beta}^{(k)}-m_{\alpha}^{(k)} \Big )^{2}} ,
\end{equation}
where the index $k$ denotes the epoch along the light curve, and the free indexes $\beta, \alpha$ can only take the above combination of values to match the free index $i$ on the right side.

The matrix $V$ is the $6 \times 6$ variance-covariance matrix:
\begin{equation}\label{eq:covariance_matrix_element_1}
V_{ij} = \langle \mathcal{R}_i^{obs} \mathcal{R}_j^{obs} \rangle = \Big \langle \sqrt{\frac{1}{N} \sum_k \Big ( m_{\beta}^{(k)}-m_{\alpha}^{(k)} \Big )^{2}} \sqrt{\frac{1}{N} \sum_p \Big ( m_{\epsilon}^{(p)}-m_{\gamma}^{(p)} \Big )^{2}} \Big \rangle.
\end{equation}

Linearization of each square root yields:
\begin{equation}\label{eq:covariance_matrix_element_2}
V_{ij} \approx \Big \langle \frac{1}{N} \sum_k \frac{(m_{\beta}^{(k)}-m_{\alpha}^{(k)})(\delta m_{\beta}^{(k)}-\delta m_{\alpha}^{(k)})}{\mathcal{R}_{\beta, \alpha}} \sum_p \frac{(m_{\epsilon}^{(p)}-m_{\gamma}^{(p)})(\delta m_{\epsilon}^{(p)}-\delta m_{\gamma}^{(p)})}{\mathcal{R}_{\epsilon, \gamma}} \Big \rangle .
\end{equation}

There can possibly be correlation for the same epoch (i.e. for $b=a$) and for those combinations of image pairs with at least one image in common, which we choose to be $\beta = \epsilon$, so this expression can be re-arranged as:
\begin{equation}\label{eq:covariance_matrix_element_final}
V_{ij} \approx \frac{1}{N} \cdot \frac{1}{\mathcal{R}_{\beta, \alpha} \cdot \mathcal{R}_{\epsilon, \gamma}} \sum_{k} (m_{\beta}-m_{\alpha}) (m_{\epsilon}-m_{\gamma})(\sigma_{\beta}^{2} \delta_{\beta \epsilon} + \sigma_{\alpha}^{2} \delta_{\beta \epsilon} \delta_{\alpha \epsilon}) .
\end{equation}
where $\sigma_{\beta}$ is the notation for the observational error in image $\beta$ and $\delta_{\beta \alpha}$ is the Kronecker symbol. We have dropped the index $k$ altogether for clarity.

Numerical implementation of this matrix element into Eq. \ref{eq:chi2_six_dimensional} allows us to evaluate the $\chi^2$  of the differences between the simulated and the observational data. We then compute the likelihood of a high number of trials as in Eq. \ref{eq:average_of_trial_likelihoods}, recovering the joint distribution as in Eq. \ref{eq:definelikelihood}.

To test this method we have also generated $200$ simulated observational campaigns, but this time each consists of a track on the maps with $30$ measures and a total length of $3$ Einstein radii, in the a set of maps corresponding to sources of sizes $0.5$, $0.1$, and $0.01$ Einstein radii for the same $4$ lenses. To each simulated epoch we have assigned an observational uncertainty of $0.2$ mag. We remark at this point that we have generated the mock observations from a different set of maps where the random seed of the star positions generator is different, resulting in fully different caustic patterns for the same physical parameters.

\begin{figure}
\epsscale{1.00}
\plotone{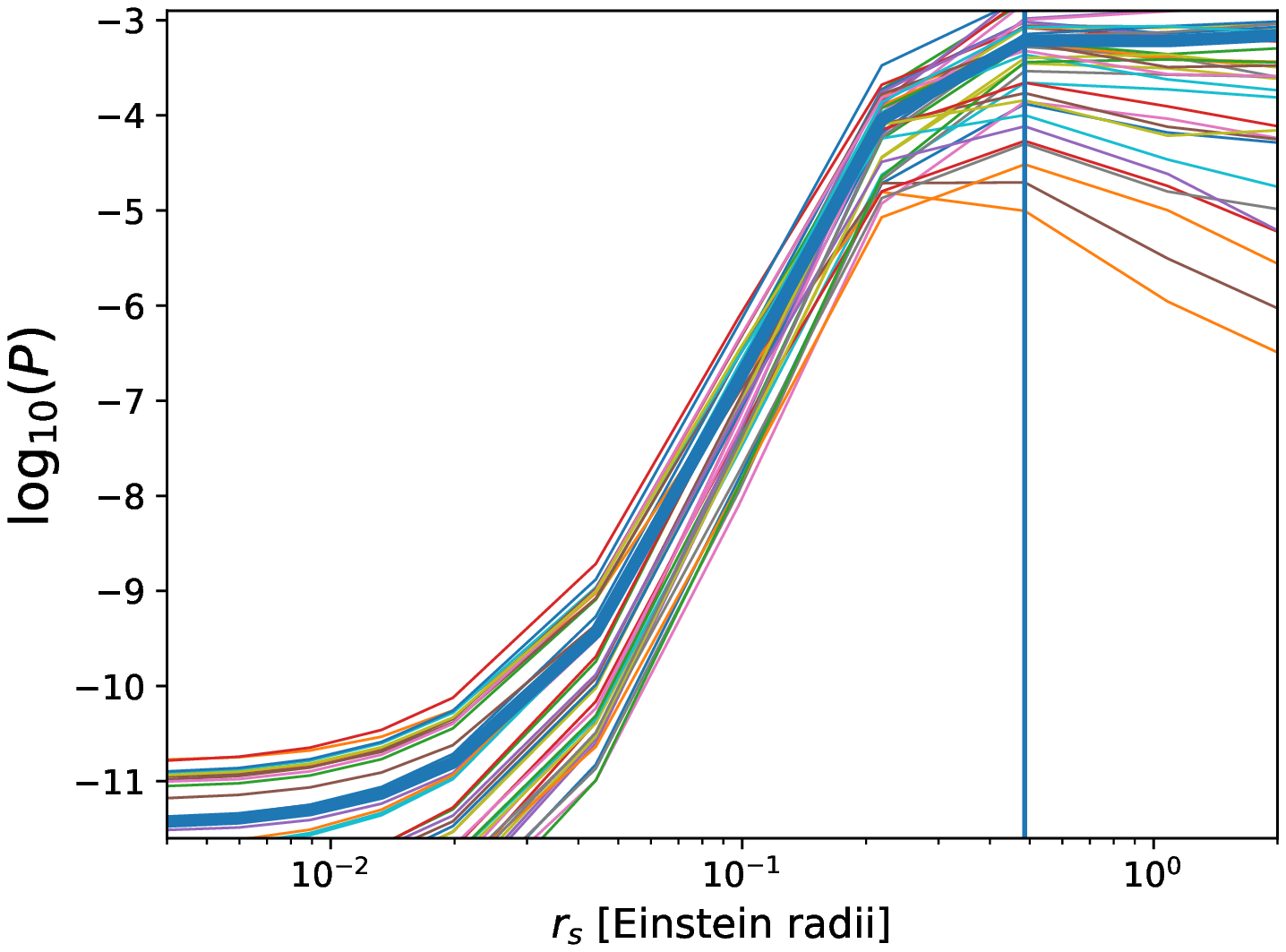}
\caption{Joint-likelihood functions (not normalized) for the source size in Einstein radii, computed from the values of the quantity $\mathcal{R}$ of $40$ simulated campaigns of $30$ observations along $3$ Einstein radii each, from magnification patterns corresponding to a source size of $0.5$ Einsteinradii (vertical blue line). The thick, blue curve represents an average. \label{fig:detail_rms_big}}
\end{figure}

\begin{figure}
\epsscale{1.00}
\plotone{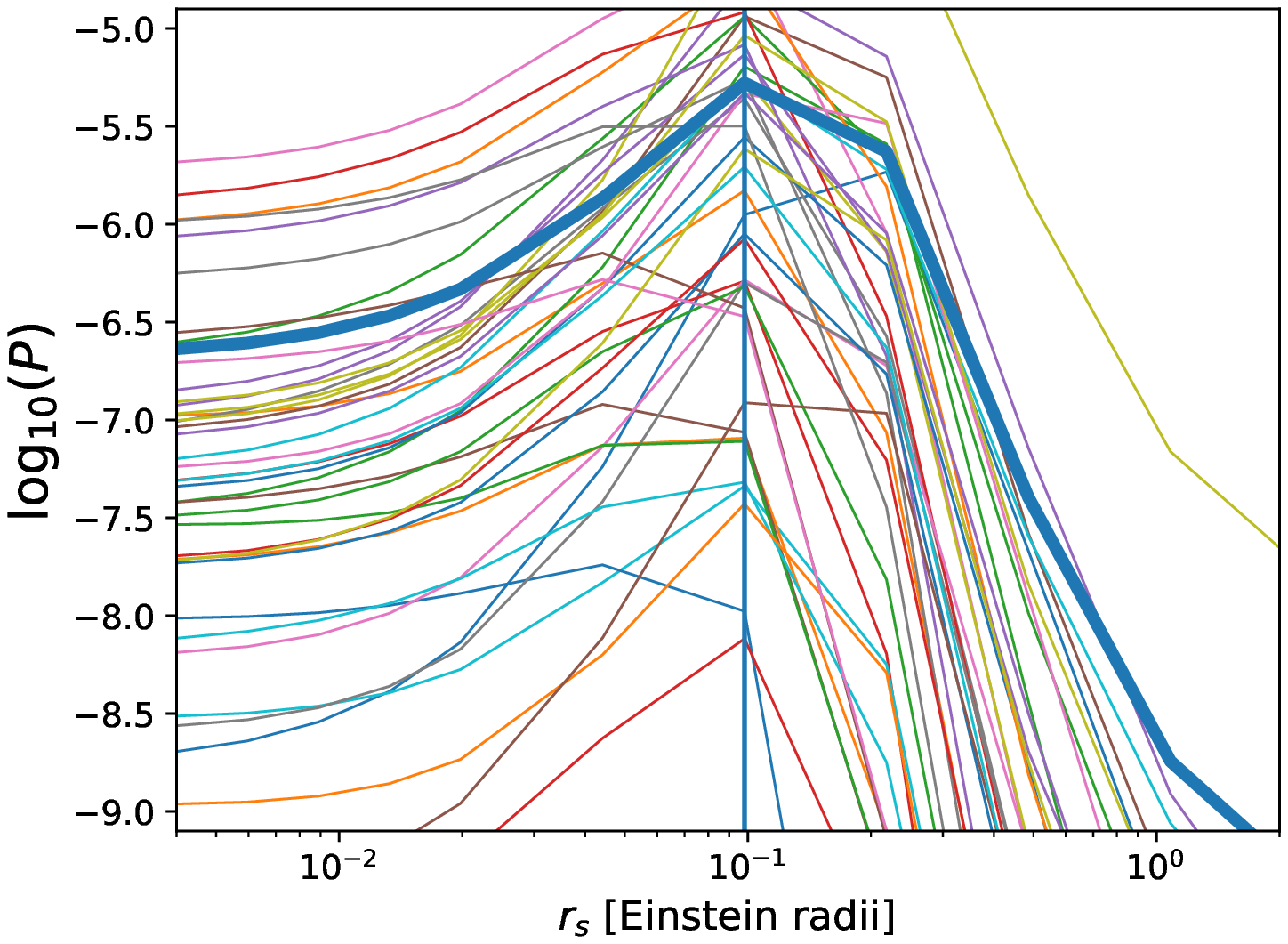}
\caption{Joint-likelihood functions (not normalized) for the source size in Einstein radii, computed from the values of the quantity $\mathcal{R}$ of $40$ simulated campaigns of $30$ observations along $3$ Einstein radii each, from magnification patterns corresponding to a source size of $0.1$ Einsteinradii (vertical blue line). The thick, blue curve represents an average. \label{fig:detail_rms_medium}}
\end{figure}

\begin{figure}
\epsscale{1.00}
\plotone{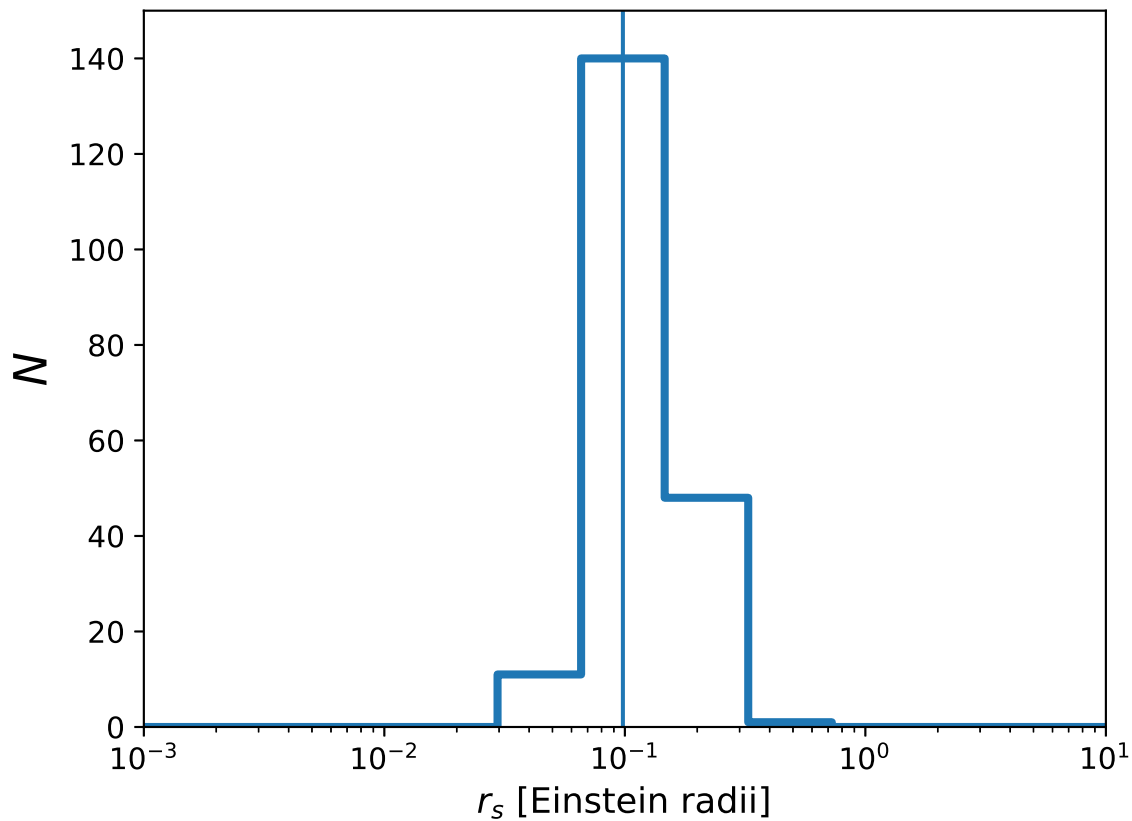}
\caption{Histogram of expected values of source size, from joint likelihood functions computed usint the values of $\mathcal{R}$ of $200$ simulated random campaigns of $30$ observations along $3$ Einstein radii each, using magnification patterns corresponding to a source size of $0.1$ Einsteinradii (vertical blue line). Note that the other figures show $40$ random trials for clarity, but this histogram is made out of $200$ trials. \label{fig:the_histogram}}
\end{figure}

\begin{figure}
\epsscale{1.00}
\plotone{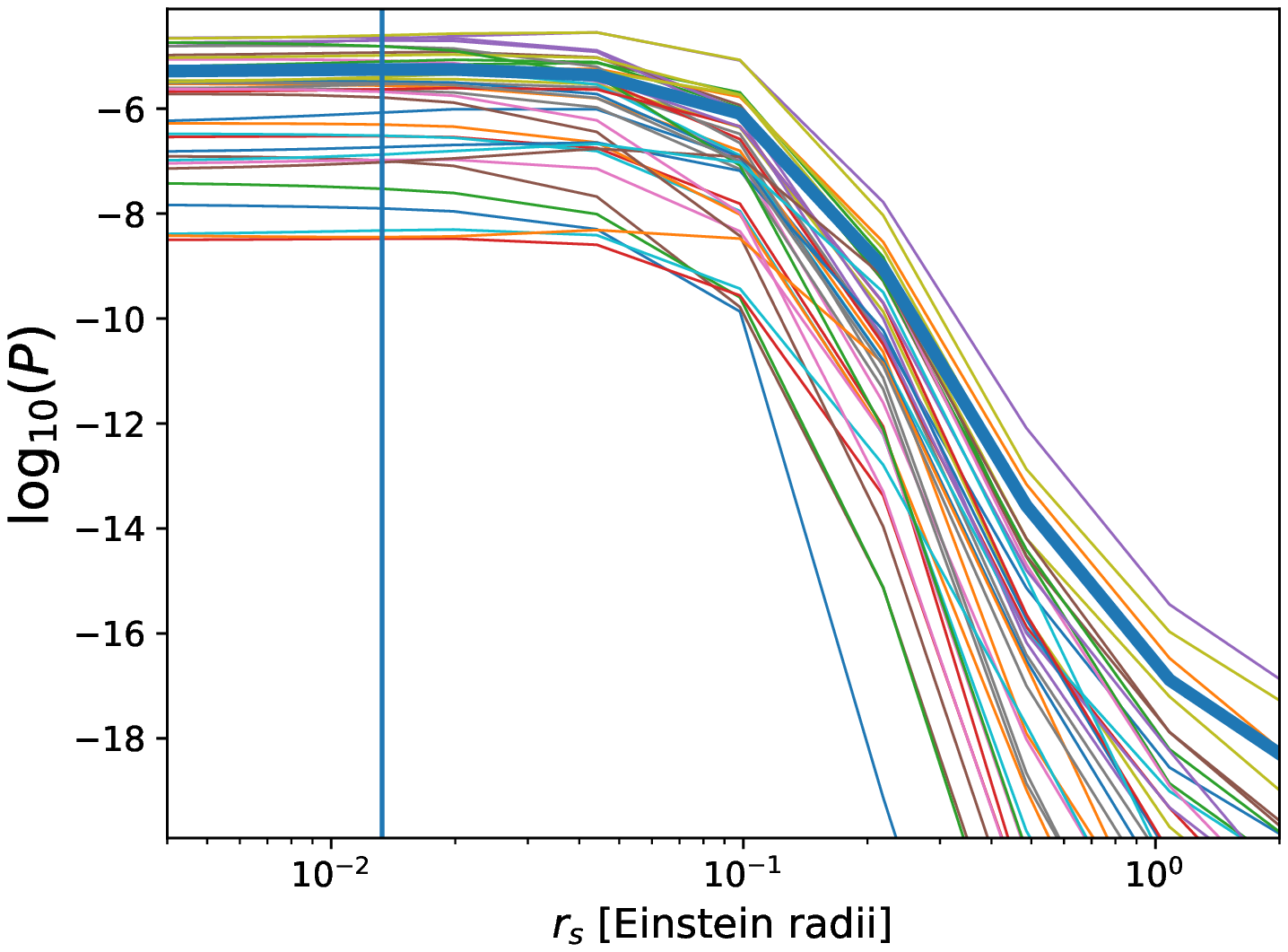}
\caption{Joint-likelihood functions (not normalized) for the source size in Einstein radii, computed from the values of the quantity $\mathcal{R}$ of $40$ simulated campaigns of $30$ observations along $3$ Einstein radii each, from magnification patterns corresponding to a source size of $0.01$ Einsteinradii (vertical blue line). The thick, blue curve represents an average. \label{fig:detail_rms_small}}
\end{figure}

The results are similar in the case of the big source of $0.5$ Einstein radii, i.e. we get a lower constraint on the size (Fig. \ref{fig:detail_rms_big}). The significant improvement is in the behavior for the small (Fig. \ref{fig:detail_rms_small}) and medium (Fig. \ref{fig:detail_rms_medium}) sources. In the case of the medium source, the expected values of each of the $200$ trials recovered individually either the exact source size, or the closest bigger value, where $140$ out of $200$ trials yielded the exact index (Fig. \ref{fig:the_histogram}) corresponding to expected values in the interval $0.07$ to $0.15$ Einstein radii. 

For the smallest source we get an upper limit consistently across all the simulations. Whereas this is a limitation of the quantity $\mathcal{R}$ or simply the result of the sampling frequency we impose (our simulated observations are $0.1$ Einstein radii apart whereas the pixel size goes down to $0.004$ Einstein radii in the simulations), it remains an open question for further studies. But most important, the method does not yield false results even in this extreme size limit.

All lengths have been expressed in units of Einstein radii. Scaling them to any specific mass value is straightforward by means of Def. \ref{eq:AE}. The application of these results therefore requires independent constraints either on the source size or the mass of the deflectors that are inducing the microlensing variability in the wavelength range of our observations.


\section{Application to real data}\label{sec:4th_section}

To show an application to real data, we have used the observational campaigns from the \textit{Chandra} archive for the same quadruple quasars, using the X-ray counts of the tables in \cite{Guerras2017}. We have scaled this problem in gravitational radii of the mass of the central SMBH (see references in the same article), using as a basis for numerical simulations magnification patterns $8000$ pixel in size with a physical scale of $1.5$ gravitational radii per pixel. This results in different sizes in light days in the lens plane, namely $342$ ld, $75$ ld, $267$ ld, and $821$ ld for HE 0435$-$1223, SDSS 0924+0219, SDSS 1004+4112 and Q 2237+0305 respectively, but as the X-ray corona is thought to be proportional to the mass of the central black hole, this scaling is a natural choice for a combined study. To simulate sources of different sizes we have convolved the maps with a Gaussian kernel $I(r) \propto \exp[-r^2/(2r_{s}^{2})]$ where $R_{s}/1.5=e^{0.2n}$ with $n= 0, 1, 2, ..., 24$, from $1.5$ Gravitational radii all the way up to $182$ Gravitational radii. For the masses, we repeat the calculations with uniform distributions of point deflectors following a sequence $10^{-n/2}$ $M_{\odot}$ with $n= 0, 1,...,4$ which spans a range of masses from $0.01$ to $1.00$ $M_{\odot}$ in order to explore the results of the analysis when different values for the deflector mass are assumed.

We then apply the procedure exposed in Sec. \ref{sec:4th_section} using the real observational campaigns for the objects, containing $10$, $6$, $11$, and $30$ epochs for HE 0435$-$1223, SDSS 0924+0219, SDSS 1004+4112 and Q 2237+0305 respectively. The tracks on the maps are given a length in gravitational radii based on the estimates of the relative transverse speed of the sources given by \cite{MosqueraKochanek2011} and the distributions of the data points along the tracks mimic the date distributions of the observations.

\begin{figure}
\epsscale{1.00}
\plotone{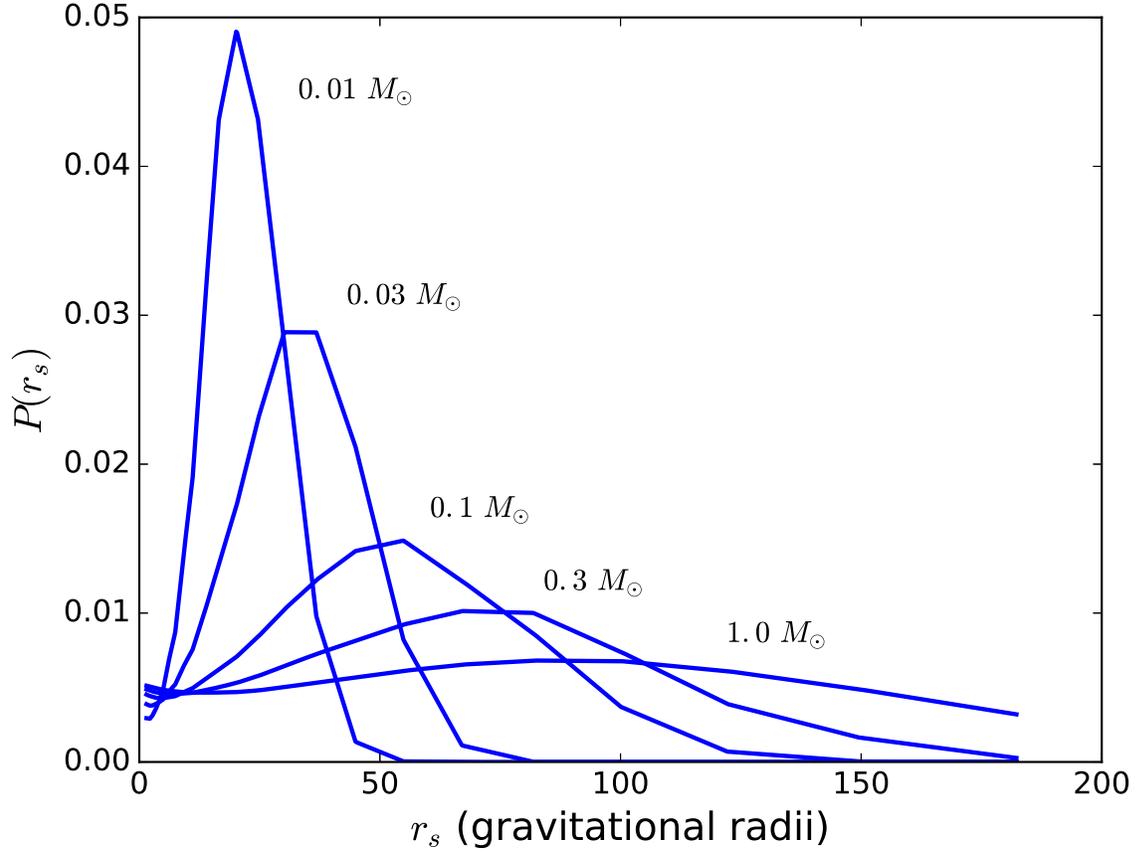}
\caption{Joint probability distribution for the X-ray emitting region size, obtained from the values of the quantity $\mathcal{R}$ of the observational campaigns of HE 0435$-$1223, SDSS 0924+0219, SDSS 1004+4112 and Q 2237+0305 assuming uniform stellar populations ranging between $0.01$ and $1.0$ $M_{\odot}$. \label{fig:the_real_joint}}
\end{figure}

\begin{figure}
\epsscale{1.00}
\plotone{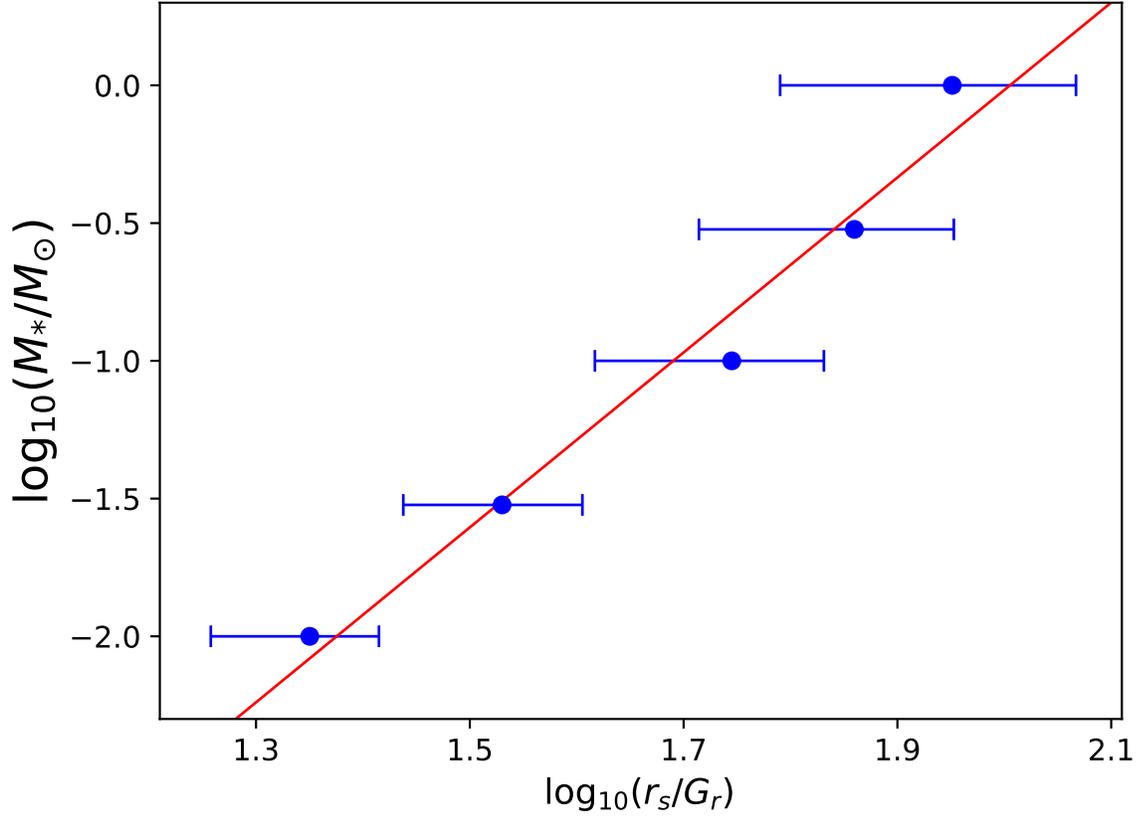}
\caption{Deflector mass vs Source size, obtained from the values of the quantity $\mathcal{R}$ of the observational campaigns of HE 0435$-$1223, SDSS 0924+0219, SDSS 1004+4112 and Q 2237+0305 assuming uniform stellar populations. The deflector mass varies, but the projected surface density is the same. Note the logarithmic scale in the axis. The red line shows a linear regression. \label{fig:degeneration}}
\end{figure}

As a result of applying the procedure described in Sec. \ref{sec:4th_section} we get a deflector mass dependent joint likelihood function for the source size based on the quantity $\mathcal{R}$ computed from the microlensing ratios of the observational campaigns. Fig. \ref{fig:the_real_joint} shows the resulting probability densities. The expected values for deflector mass $0.01$, $0.03$, $0.1$, $0.3$, and $1.0$ $M_{\odot}$ are $22.4^{+3.6}_{-4.3}$, $33.9^{+6.4}_{-6.5}$, $55.6^{+12.2}_{-14.2}$, $72.4^{+17.3}_{-20.6}$, and  $89.4^{+27.3}_{-27.7}$ gravitational radii respectively (one-sigma intervals). This shows a degeneracy between deflector mass and source size that is best understood in Fig. \ref{fig:degeneration}, where a red line shows the linear regression:
\begin{equation}\label{eq:linear_regression}
\log_{10}(M/M_{\odot}) = (3.2 \pm 0.3) \log_{10}(r_g) - (6.4 \pm 0.5)
\end{equation}

Similarly, a degeneracy has been noticed in the track fitting method \citep{Kochanek2004}, where the source size determination depends either on the mean deflector mass or the speed of the deflectors. Combined with a velocity prior, the track fitting method is able to break this degeneracy and measure source sizes \citep[e.g.,][]{Mosquera2013,Blackburne2015,Macleod2015}. However, the velocity information is lost in the $\mathcal{R}$ metric, and independent constraints either on source size or mean deflector mass are required. If we combine the source size values for the X-ray corona found in other joint studies with Eq. \ref{eq:linear_regression}, we can obtain at this point an estimate of the mass of the deflectors that produce X-ray variability. \cite{Jorge2015} found $r_{1/2} \simeq (24 \pm 14)$ gravitational radii in a combined study of $10$ lensed quasars (half-light radius), and \cite{Guerras2017} found $r_{1/2} \simeq (14 \pm 11)$ gravitational radii in a sample with $6$ lenses. Those values correspond to mass intervals spanning from $6 \times 10^{-3}$ and and $1 \times 10^{-3}$ $M_{\odot}$ respectively (upper 1$\sigma$ limits) down to zero, according to the source size-deflector mass degeneracy of Eq. \ref{eq:linear_regression}. The conclusion would be that X-ray microlensing is mostly due to planetary mass objects, in agreement with recent results based on the anomalous shift of the $K_{\alpha}$ iron line \citep{DaiGuerras2018}. However there are some aspects that deserve further discussion before we can embrace that conclusion.

The process of computing magnification patterns for increasingly small deflector mass and comparing with the observational data cannot proceed indefinitely, because the sampling frequency \footnote{The time intervals between observations is variable in our sample, but assuming a sampling frequency is a useful concept for discussing qualitative aspects here.} in the data is fixed, whereas the Einstein radius (and the typical microlensing timescales) goes as $\sqrt{M_{*}}$. The data for the objects in our sample are separated in average by $0.8$, $1.1$, $0.9$ and $0.5$ years for HE 0435$-$1223, SDSS 0924+0219, SDSS 1004+4112 and Q 2237+0305 respectively \citep[see tables in][]{Guerras2017} with some intervals better resolved, whereas the caustic crossing timescales as estimated by \cite{MosqueraKochanek2011} are $0.47$, $0.39$, $2.2$, and $0.23$ $\sqrt{M_{*}/(0.3 M_{\odot})}$, which certainly sets a physical limitation when we try to work with very low mass populations. It can be argued that in our method, by construction, the simulated patterns in the lens plane are undersampled in the same amount as the real observations, but there is also a limit for this. As with any other extrapolation, caution is advised in interpreting the results.


\section{Conclusion}\label{sec:Conclusion}

We have studied some aspects in which the square root of the mean quadratic distance to the baseline value in microlensing curves, can be useful in the study of lensed quasars. We have seen that this quantity carries information about the system, and that this information is kept even for very long observational campaigns.

We have designed and put to test a simple method to recover this information in simulated scenarios, finding that the new quantity is useful to recover a source size around $0.1$ Einstein radii, which can be easily scaled to any mass range. In the simulations we also tried to force the detection down to a hypothetical source size of $0.01$ Einstein radii, and we were able to get an upper limit. This limit may have to do with the sampling frequency of our simulations. On the other side of the scale, experimenting with typical sizes above $10$ Einstein radii in our simulations yielded a lower bound. In neither one or the other extremes the method shows limitation in which it yields only one-sided constraints, but in neither case we got false results.

The application of this method to a limited sample of $4$ quasars from the \textit{Chandra} X-ray archive yielded an average size of $71 \pm 19$ gravitational radii under the initial assumption of a uniform stellar populations of $0.3$ $M_{\odot}$. This is in good agreement only with the estimate given by \cite{Mosquera2013} for the hard X-ray region in Q 2237+0305 (between $71$ and $84$ gravitational radii, but they also give $32$ to $36$ gravitational radii for the soft band). But it is a larger value than the estimates found for other objects, e.g $10$ gravitational or less for HE 0435$-$1223 \citep{Blackburne2015} or the upper limit of $62$ gravitational radii given by \cite{Macleod2015} for SDSS 0924+0219. The average sizes found in joint studies are also smaller, e.g. \cite{Jorge2015} who found an average size of $24 \pm 14$ gravitational radii based on a single-epoch approach, or our previous result \citep{Guerras2017} of $14 \pm 11$ gravitational radii based on a numerical treatment of the averages of microlensing ratios in a smaller sample. Since our simulations have recovered the source size correctly this disparity between our results and the previous estimates shows that the value of $0.3$ $M_{\odot}$ for the deflectors may not be appropriate here. In successive attempts with smaller deflector masses the resulting sizes are smaller, down to $22 \pm 4$ gravitational radii for $0.01$ $M_{\odot}$ deflectors. This shows an inverse relation between deflector mass and source size that requires using this method together with additional constraints to one or the other parameter. An extrapolation of this relation suggests that X-ray microlensing is produced by planetary mass objects, in agreement with the recent study by \cite{DaiGuerras2018} based on the anomalous shift of the $K_{\alpha}$ iron. This would also explain the lack of correlation between the microlensing variability in optical and X-ray observations found by \cite{Mosquera2013}, in which the former could be due to stars and the later produced by a scattered population of nanolenses of planetary mass.

A certain number of simplifying assumptions have been useful in this first study, and they deserve further attention in the next future. We have ignored the time-delay in the simulations, although we have seen that by increasing the duration of the campaigns, the new observable tends to stabilize. In other words, when the length of the observational campaigns is noticeably greater than the time delay, it is expected that the influence of this will be negligible. Also, we have used uniform stellar populations, but a more reallistic approach may require a model of evolved stellar populations with several components.

Finally, we note that the results in the section \ref{sec:4th_section} do not invalidate the single-epoch approach, but point to the need to include a sufficient number of objects in the sample to stabilize the results. If there is any conclusion to this part of our study, it is that four objects is too small a sample, even being lenses with $4$ images, and that the results of single-epoch studies on small samples probably overestimate the source size.

The simulations we have done cover very long campaigns, where averaging curves is meaningless due to the loss of information. The method we have outlined here is computationally cheap and can be helpful when characterizing a set of objects, for example radio-loud vs. radio-quiet or any other classification that we want to do on a database of a large number of objects, a situation that awaits us in a decade from now, given the imminent entry into operation of a new generation of space observatories.


\section{Acknowledgements}
We thank Christopher Kochanek for his comments that greatly helped shaping Sec. \ref{subsec:generalization}. Support for this work was provided by the National Aeronautics and Space Administration through {\emph{Chandra}} Award Number GO0-11121, GO1-12139, GO2-13132, and G03-14110 issued by the {\emph{Chandra}} X-ray Observatory Center, which is operated by the Smithsonian Astrophysical Observatory for and on behalf of the National Aeronautics Space Administration under contract NAS8-03060. XD acknowledges NASA ADAP program NNX15AF04G, NNX17AF26G and NSF grant AST–1413056.


\clearpage


\begin{figure}
\epsscale{1.00}
\plotone{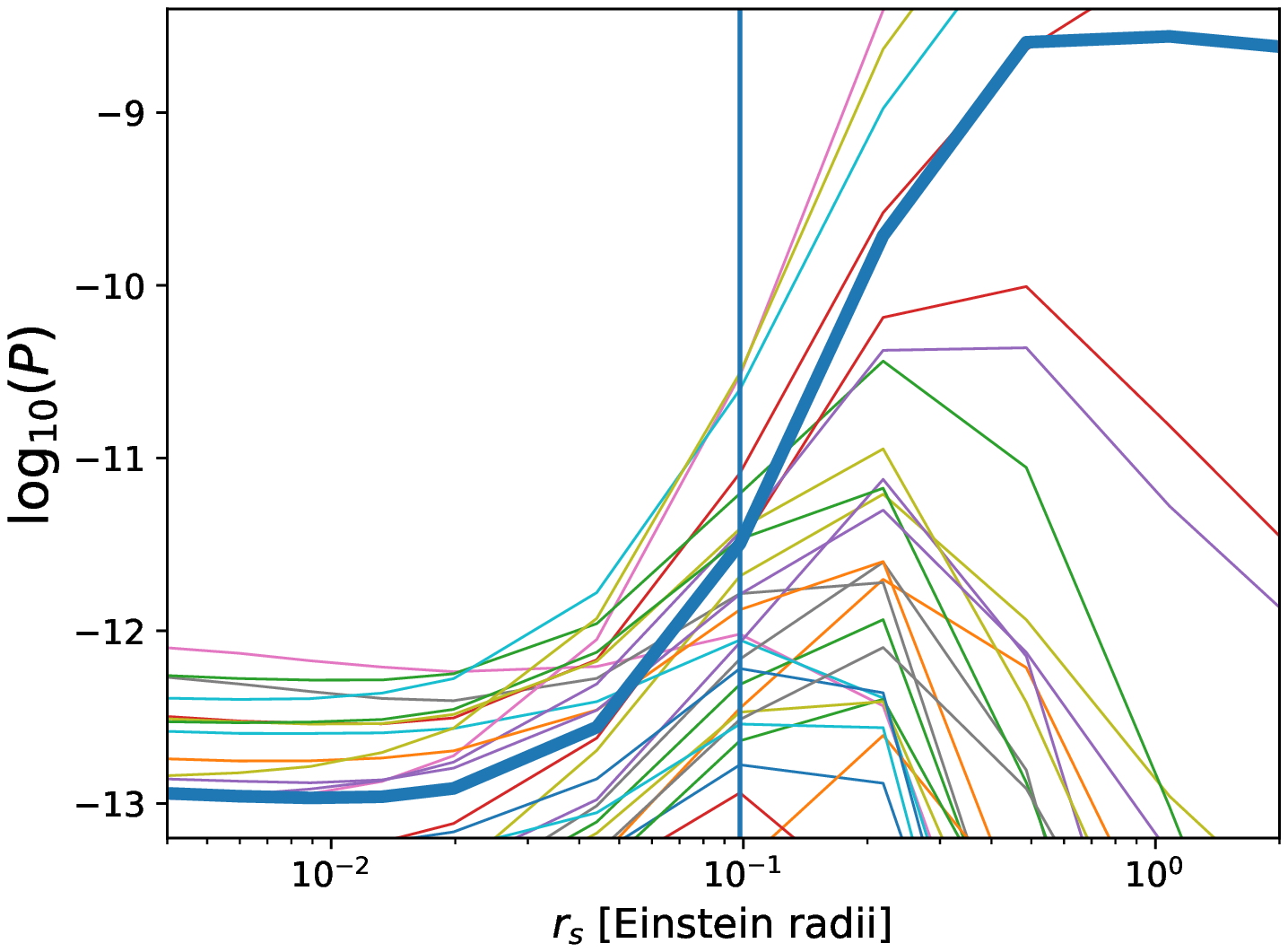}
\caption{Single-epoch joint-likelihood functions (not normalized) for $40$ random single-epoch observations simulated from magnification patterns corresponding to a source size of $0.1$ Einsteinradii (vertical blue line). The thick, blue curve represents an average. \label{fig:detail_single_medium}}
\end{figure}

\begin{figure}
\epsscale{1.00}
\plotone{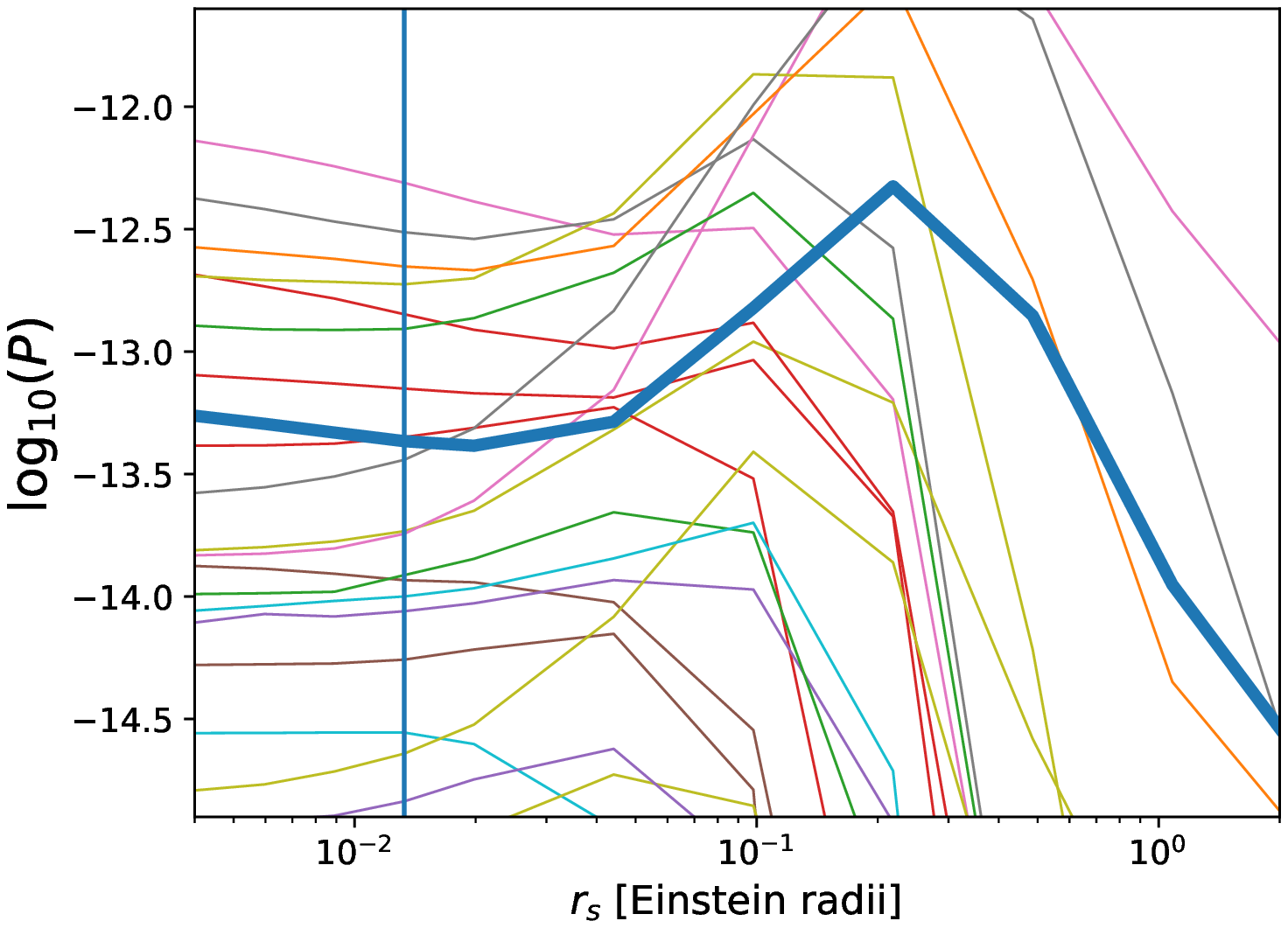}
\caption{Single-epoch joint-likelihood functions (not normalized) for $40$ random single-epoch observations simulated from magnification patterns corresponding to a source size of $0.01$ Einstein radii (vertical blue line). The thick, blue curve represents an average. \label{fig:detail_single_small}}
\end{figure}

\begin{deluxetable}{lllllll}
\tablewidth{0pt}
\tabletypesize{\scriptsize}
\tablecolumns{4}

\centering

\tablecaption{Lens model properties at the images \label{tab:kappasgammas}}

\tablehead{
\colhead{Object} &
\colhead{Image} &
\colhead{$R/R_{ef}$} &
\colhead{$\kappa_{*} / \kappa$ } &
\colhead{$\kappa$ } &
\colhead{$\gamma$ } &
\colhead{Macrolens model}
}

\startdata

HE 0435$-$1223  &  A  &  1.71  &  0.27  &  0.445  &  0.383  &  SIE+g  \\
HE 0435$-$1223  &  B  &  1.54  &  0.31  &  0.539  &  0.602  &  SIE+g  \\
HE 0435$-$1223  &  C  &  1.71  &  0.27  &  0.444  &  0.396  &  SIE+g  \\
HE 0435$-$1223  &  D  &  1.40  &  0.34  &  0.587  &  0.648  &  SIE+g  \\
SDSS 0924+0219  &  A  &  2.93  &  0.12  &  0.472  &  0.456  &  SIE+g  \\
SDSS 0924+0219  &  B  &  3.26  &  0.10  &  0.443  &  0.383  &  SIE+g  \\
SDSS 0924+0219  &  C  &  2.69  &  0.14  &  0.570  &  0.591  &  SIE+g  \\
SDSS 0924+0219  &  D  &  2.79  &  0.13  &  0.506  &  0.568  &  SIE+g  \\
SDSS 1004+4112  &  A  &  $-$   &  0.03  &  0.763  &  0.300  &  parametric    \\
SDSS 1004+4112  &  B  &  $-$   &  0.03  &  0.696  &  0.204  &  parametric    \\
SDSS 1004+4112  &  C  &  $-$   &  0.03  &  0.635  &  0.218  &  parametric    \\
SDSS 1004+4112  &  D  &  $-$   &  0.03  &  0.943  &  0.421  &  parametric    \\
Q 2237+0305     &  A  &  0.24  &  0.79  &  0.39   &  0.40   &  SIE+g  \\
Q 2237+0305     &  B  &  0.25  &  0.79  &  0.38   &  0.39   &  SIE+g  \\
Q 2237+0305     &  C  &  0.20  &  0.81  &  0.74   &  0.73   &  SIE+g  \\
Q 2237+0305     &  D  &  0.23  &  0.80  &  0.64   &  0.62   &  SIE+g  \\
\enddata

\tablecomments{For each image we give the distance $R/R_{ef}$ of the image from the lens center in units of the effective radius of the lens from \cite{Oguri2014}, the expected fraction of the surface density in stars $\kappa_{*}/\kappa$, the surface density in stars, the surface density $\kappa$ in units of the lens critical density and the total shear $\gamma$. We used the models of \cite{Schechter2014} for HE 0435$-$1223 and SDSS 0924+0219, \cite{Kochanek2004} for Q 2237+0305, and \cite{Oguri2014} for SDSS 1004+4112}

\end{deluxetable}


\clearpage

\begin{deluxetable}{lccccccccc}
\tabletypesize{\scriptsize}

\tablecaption{Lens Data\label{tab:summaryofdata}}
\tablewidth{0pt}

\tablehead{
\colhead{Object} & \colhead{$z_{s}$} & \colhead{$z_{l}$} & \multicolumn{1}{c}{$R_{E}$} & \colhead{Epochs} & \colhead{$t_E$} & \colhead{$t_s$} & \colhead{$\Delta t_{obs}$} &  \colhead{$\Delta t_{obs}/t_E$}  &\colhead{$M_{BH}$}         \\
                 &                   &                   & (light-days)                & (years)         & (years)         & (years)                    &  (Einstein radii)                &($\times 10^{9}$ $M_{\odot}$)
}

\startdata

HE 0435$-$1223  &  1.689  &  0.46    &  7.986 & 10 & 18.3 & 0.47  &  7.3   &  0.40  &  $0.50$ $ $  $ $(CIV) \\
SDSS 0924+0219  &  1.524  &  0.39    &  7.790 &  6 & 20.4 & 0.39  &  5.6   &  0.27  &  $0.11$ $ $  $ $(MgII)\\
SDSS 1004+4112  &  1.734  &  0.68    &  7.737 & 11 & 28.9 & 0.28  &  9.4   &  0.33  &  $0.39$ $ $  $ $(MgII)\\
Q 2237+0305     &  1.69   &  0.0395  &  3.660 & 30 & 8.11 & 0.23  &  13.6  &  1.68  &  $1.20$ $ $  $ $(H$_{\beta}$) 

\enddata

\tablecomments{Based on the source and lens redshifts $z_{s}$ and $z_{l}$, the Einstein radius $R_{E}$ can be computed. Here we report the estimates given by \cite{MosqueraKochanek2011} of $R_{E}$ as well as the Einstein radius and source crossing time scales $t_E$ and $t_s$, assuming a mean stellar mass in lens galaxies of $\langle M_{*} \rangle = 0.3$ $ M_{\odot}$, for comparison to the time span of the observations $\Delta t_{obs}$. The last column reports the estimated black hole mass and the emission lines used for the estimates by \cite{Morgan2010} (HE 0435$-$1223, SDSS 1004+4112)\cite{Peng2006} (SDSS 0924+0219) and \cite{Assef2011} (Q 2237+0305).}

\end{deluxetable}

%


\begin{thebibliography}{}

\bibitem[Assef et al.(2011)]{Assef2011}
Assef, R. J., Denney, K. D., Kochanek, C. S., Peterson, B. M., et al. 2011, \apj, 742, 93 

\bibitem[Blackburne et al.(2014)]{Blackburne2014}
Blackburne, J. A., Kochanek, C. S., Chen, B., Dai, X., \& Chartas, G. 2014, \apj, 789, 125 
  
\bibitem[Blackburne et al.(2015)]{Blackburne2015}
Blackburne, J. A., Kochanek, C. S., Chen, B., Dai, X., \& Chartas, G. 2015, \apj, 798, 95 

\bibitem[Chen et al.(2011)]{Chen2011} Chen, B., Dai, X., Kochanek, C.~S., et al.\ 2011, \apjl, 740, L34

\bibitem[Chen et al.(2012)]{Chen2012}
Chen, B., Dai, X., Kochanek, C. S., Chartas, G., et al. 2012, \apj, 755, 24 

\bibitem[Dai et al.(2003)]{Dai2003}
Dai, X., Chartas, G., Agol, E., Bautz, M. W., \& Garmire, G. P. 2003, \apj, 589, 100 

\bibitem[Dai et al.(2010)]{Dai2010}
Dai, X., Kochanek, C. S., Chartas, G., Kozłowski, S., et al. 2010, \apj, 709, 278 

\bibitem[Dai \& Guerras(2018)]{DaiGuerras2018}
Dai, X., \& Guerras, E. 2018, \apj, 853, 27 

\bibitem[Fian et al.(2018)]{Fian2018}
Fian, C., Guerras, E., Mediavilla, E., Jim\'enez-Vicente, J., et al. 2018, \apj, 859, 50 

\bibitem[Guerras et al.(2017)]{Guerras2017}
Guerras, E., Dai, X., Steele, S., Ang, L., et al. 2017, \apj, 836, 206 

\bibitem[Guerras et al.(2013a)]{Guerras2013a}
Guerras, E., Mediavilla, E., Jim\'enez-Vicente, J., Kochanek, C. S., et al. 2013, \apj, 764, 160 

\bibitem[Guerras et al.(2013b)]{Guerras2013b}
Guerras, E., Mediavilla, E., Jim\'enez-Vicente, J., Kochanek, C. S., et al. 2013,. \apj, 778, 123 

\bibitem[Jim\'enez-Vicente et al.(2015)]{Jorge2015}
Jim\'enez-Vicente, J., Mediavilla, E., Kochanek, C. S., \& Mu\~noz, J. A. 2015, \apj, 806, 251 

\bibitem[Kochanek(2004)]{Kochanek2004}
Kochanek, C. S. 2004, \apj, 605, 58 

\bibitem[MacLeod et al.(2015)]{Macleod2015}
MacLeod, C. L., Morgan, C. W., Mosquera, A., Kochanek, C. S., et al. 2015, \apj, 806, 258 

\bibitem[Mediavilla et al.(2015)]{Mediavilla2015} 
Mediavilla, E., Jim\'enez-Vicente, J., Mu\~noz, J. A., Mediavilla, T., \& Ariza, O. 2015, \apj, 798, 138 

\bibitem[Mediavilla et al.(2011a)]{Mediavilla2011a}
Mediavilla, E., Mu\~noz, J. A., Kochanek, C. S., Guerras, E., et al. 2011, \apj, 730, 16 

\bibitem[Mediavilla et al.(2011b)]{Mediavilla2011b}
Mediavilla, E., Mediavilla, T., Mu\~noz, J. A., Ariza, O., et al. 2011, \apj, 741, 42 

\bibitem[Mediavilla et al.(2006)]{Mediavilla2006}
Mediavilla, E., Mu\~noz, J. A., Lopez, P., Mediavilla, T., et al. 2006, \apj, 653, 942 

\bibitem[Morgan et al.(2008)]{morgan2008} Morgan, C.~W., Kochanek, C.~S., Dai, X., Morgan, N.~D., \& Falco, E.~E.\ 2008, \apj, 689, 755-761

\bibitem[Morgan et al.(2010)]{Morgan2010}
Morgan, C. W., Kochanek, C. S., Morgan, N. D., \& Falco, E. E. 2010, \apj, 712, 1129 

\bibitem[Morgan et al.(2012)]{Morgan2012}
Morgan, C. W., Hainline, L. J., Chen, B., Tewes, M., et al. 2012, \apj, 756, 52 

\bibitem[Mosquera \& Kochanek(2011)]{MosqueraKochanek2011}
Mosquera, M., \& Kochanek, C. S. 2011, \apj, 738, 96 

\bibitem[Mosquera et al.(2013)]{Mosquera2013}
    Mosquera, A. M., Kochanek, C. S., Chen, B., Dai, X., Blackburne, \& J. A., Chartas, G. 2013, \apj, 769, 53 

\bibitem[Mu\~noz et al.(2016)]{Munoz2016}
Mu\~noz, J. A., Vives-Arias, H., Mosquera, A. M., Jiménez-Vicente, J., et al. 2016, \apj, 817, 155 

\bibitem[Oguri et al.(2014)]{Oguri2014}
Oguri, M., Rusu, C. E., \& Falco, E. E. 2014, \mnras, 439, 2494 

\bibitem[Peng et al.(2006)]{Peng2006}
Peng, C. Y., Impey, C. D., Rix, H. W., Kochanek, C. S., et al. 2006, \apj, 649, 616 

\bibitem[Schechter et al.(2004)]{Schechter2004}
Schechter, P. L., Wambsganss, J., \& Lewis, G. F. 2004, \apj, 613, 77 

\bibitem[Schechter et al.(2014)]{Schechter2014}
Schechter, P. L., Pooley, D., Blackburne, J. A., \& Wambsganss, J. 2014, \apj, 793, 96 

\bibitem[Vernardos \& Fluke(2013)]{Vernardos2013}
Vernardos, G., \& Fluke, C. J. 2013, \mnras, 434,  832 
 
\bibitem[Wambsganss(1992)]{Wambsganss1992}
Wambsganss, J. 1992, \apj, 386, 19 

\bibitem[Wambsganss(2006)]{Wambsganss2006}
Wambsganss, J. 2006, Saas-Fee Advanced Course 33: Gravitational Lensing: Strong, Weak and Micro, Springer-Verlag Berlin Heidelberg, ISBN 978-3-540-30309-1

\end{thebibliography}
\end{document}